%% file: standardizing.tex
\documentclass{amsart}
\usepackage{amscd}     
\usepackage{amssymb}
\usepackage{verbatim}

\theoremstyle{plain}
\newtheorem{Thm}{Theorem}

\theoremstyle{definition}
\newtheorem{Def}{Definition}
\newtheorem{Exe}{Exercise}

\theoremstyle{remark}

\input{timemacros} 
\renewcommand{\rm}{\normalshape}

\numberwithin{equation}{section}  

\begin{document}


\title[Standardizing distance and time]%
   {Standardizing distance and time}
\author{Russell Clark Eskew}
\address{PO Box 8150, Austin, Texas 78713-8150, USA}
\email{eskewr@@io.com}


\maketitle
\begin{abstract}
   Einstein's Equivalence Principle is used with the electromagnetic spectrum 
   to translate meters and seconds  
   into radians and seconds.
   Based on a unique geometric relationship,
   a new transformation of velocities and
   a changed Lorentz transformation result.
   The physical angle of parallelism is quantified.
   The way we measure distance and time is standardized,
   constructing a theory of time and geometry to the universe.
\end{abstract}

\newpage

\section{INTRODUCTION} 
In \( Speculations \) \( in \) \( Science \) \( and \) \( Technology \) {\bf 21,} \( 213-225 (1999), \)
``Time, gravity and the exterior angle of parallelism,'' 
(available on-line from http://www.wkap.nl)
the author, Russell Clark Eskew, set out to standardize distance and time 
with the electromagnetic spectrum \cite[p. 220]{rE99}.
The number of seconds to a light year illustrated how an infinite number
of seconds derive a mathematically convenient scale for the physical
angle of parallelism, \( 2 \tan^{-1} e^{-a/s} = 2 \tan^{-1} e^{-1}. \)
In this paper, we follow up by showing how the wavespeed 
of light, 299792458 meters per second,
with wavelengths in meters and radians per cycle,
apply to a frequency \( f \) in cycles per second to create
a wavespeed in \( radians \) per second
\begin{equation}\label{E:firstsec1eq}
\lambda = \tan \frac{\psi}{2} = \tanh \frac{a}{2},
\end{equation}
where \( \psi = \tan^{-1} \sinh a \) \cite[pp. 312--313]{gC31} 
\cite[p. 215]{rE99}, \( 0 < \psi < \frac{\pi}{2}, \) \( a = \ln f, \)
and \( s \) is the hyperbolic \( radius \) \( of \) \( curvature. \) 
The significance of \( \lambda = \tan \frac{\psi}{2}, \)
as Roger Penrose points out \cite[pp. 74--75]{wR69},
is that the stereographic projection of a line drawn
with an angle \( \frac{\psi}{2} \)
from the west pole of a sphere, \( P, \)
and an angle \( \psi \) from the origin,
maps the point \( Q \) onto the plane \( x = 1 \) at \( \tan \frac{\psi}{2}. \)
This preserves measuring great circles and angles with a partial metric \cite[pp. 92--95]{hC61}.

\section{A UNIQUE GEOMETRIC RELATIONSHIP}
The geometric relationship is unique in this scheme.
With \( \lambda \) we know that the hypotenuse line from the unit circle 
to the horizontal unit hyperbola, \( x^2 - y^2 = 1, \) is
\begin{equation} \label{E:fifteen}
(x^2 + y^2)^{1/2} = ((\cosh \frac{a}{2})^2 + (\sinh \frac{a}{2})^2)^{1/2} = (\frac{(e^a +1)^2 +(e^a - 1)^2}{4e^a})^{1/2} = (\frac{1 + \lambda^2}{1 - \lambda^2})^{1/2}.
\end{equation}   
We use the ratio of both hypotenuses to multiply the circular coordinates
\begin{align}
\cos \frac{\psi}{2} &= (\frac{1}{1 +\lambda^2})^{1/2} = \frac{1}{\cosh \sinh^{-1} \lambda} = \frac{1}{e^{\sinh^{-1} \lambda} - \lambda} = e^{i\psi/2} - i\sin \frac{\psi}{2} \\
\sin \frac{\psi}{2} &= (\frac{\lambda^2}{1 + \lambda^2})^{1/2} = \tanh \sinh^{-1} \lambda = \frac{\lambda}{e^{\sinh^{-1} \lambda} - \lambda} = \frac{e^{i\psi/2} - \cos \frac{\psi}{2}}{i} \notag
\end{align}
by \eqref{E:fifteen} to derive the hyperbolic coordinates
\begin{align}
\cosh \frac{a}{2} &= (\frac{1}{1 - \lambda^2})^{1/2} = \frac{1}{\cos \sin^{-1} \lambda} = \frac{1}{e^{i\sin^{-1} \lambda} - i\lambda} = e^{a/2} - \sinh \frac{a}{2} \\
\sinh \frac{a}{2} &= (\frac{\lambda^2}{1 - \lambda^2})^{1/2} = \tan \sin^{-1} \lambda = \frac{\lambda}{e^{i\sin^{-1} \lambda} - i\lambda} = e^{a/2} - \cosh \frac{a}{2}. \notag
\end{align}

\newpage
\section{A NEW TRANSFORMATION OF VELOCITIES}
This paper's transformation of velocities is different from that of Galileo Galilei (1564--1642)
\begin{equation} \label{E:firstsec3eq}
u = u' + v.
\end{equation}
The quantities are the ``absolute velocity'' \( u, \) \( i.e., \)
a moving particle's velocity with respect to a fixed reference frame,
its ``relative velocity'' \( u', \) \( i.e., \)
the particle's velocity with respect to a moving reference frame, and
the ``transport velocity'' \( v, \) \( i.e., \)
the velocity of the moving reference frame.  
Albert Einstein (1879--1955) believes when \( u' = c, \)
\( c \) being the speed of light, that \( u = c \)
rather than \( u = c + v \) \cite[pp. 161, 173, 203--212]{iI69} \cite{hL23} \cite{aE57}.
The transformation of velocities of Einstein and H. A. Lorentz (1853--1928) 
replaced \eqref{E:firstsec3eq} with 
\begin{equation}
u = \frac{u' +v}{1 + u'v/c^2}.
\end{equation}
This paper's transformation of velocities is different from that of Einstein, too.

For the derivatives of the wavespeed \( \lambda \) 
\begin{align}
\frac{d\lambda}{d\psi} &= \frac{d}{d\psi} \tan \frac{\psi}{2} = \frac{1}{2} \sec^2 \frac{\psi}{2} \\
   \frac{d\lambda}{da} &= \frac{d}{da} \tanh \frac{a}{2} = \frac{1}{2} \operatorname{sech}^2 \frac{a}{2} = \frac{2f}{(f + 1)^2} \notag
\end{align}
make this paper's transformation of velocities
\begin{align}
 \frac{d\lambda}{d\psi} &= \frac{d\lambda}{da} + \lambda^2 \\
\mathrm{acceleration_{absolute}} &= \mathrm{acceleration_{relative} + acceleration_{transport}}. \notag
\end{align}
If the wavespeed in meters, \( c, \) is a constant, then \( dc = 0. \)
But since the wavespeed in radians, \( \lambda, \) is a variable, then we might 
have \( d\lambda \neq 0. \)  
This is why we hereby replace \( c \) with \( \lambda, \) 
and \( v \) with \( \lambda^2, \) which is also known as 
the \( \mathrm{acceleration_{frame}} \) of the gravitational frame force
\( F_{\mathrm{frame}} = \mathrm{-mass \times acceleration_{frame}}, \) 
in a rotating frame of reference
\begin{align}
               F_{\psi} &= F_{a} - F_{\mathrm{frame}} \\
m\frac{d\lambda}{d\psi} &= m\frac{d\lambda}{da} + m\frac{\lambda^2}{r}. \notag
\end{align}

\section{CHANGING THE LORENTZ TRANSFORMATION}
With the concept of \( proper \) \( time, \) a moving particle with instantaneous
velocity (\( i.e., \) ``transport acceleration'' of a moving reference frame)
 \( v(t) = \lambda^2(t) \) relative to some inertial system \( K \)
(\( i.e., \) with ``absolute acceleration'' \( d\lambda/d\psi \) 
with respect to a fixed reference frame)
changes its position in a time interval \( dt \) by \( dx = v dt = \lambda^2 dt. \)
The space and time coordinates in \( K', \)
\( (t',z',x',y') = (x_0^{'},x_1^{'},x_2^{'},x_3^{'}) = (ct',z',x',y'), \)
where the system is instantaneously at rest 
(\( i.e., \) with the particle's ``relative acceleration'' \( d\lambda/da \)
with respect to the moving reference frame),
are related to those in \( K, \)
\( (t,z,x,y) = (x_0,x_1,x_2,x_3) = (ct,z,x,y), \)
by the inverse \( Lorentz \) \( transformation \)  
\begin{align}
x_0 &= \gamma(x_0^{'} + x_1^{'}\beta) = (\cosh a)(x_0^{'} + x_1^{'}\tanh a) \\
x_1 &= \gamma(x_1^{'} + x_0^{'}\beta) = (\cosh a)(x_1^{'} + x_0^{'}\tanh a) \notag \\
x_2 &= x_2^{'} \notag \\
x_3 &= x_3^{'}. \notag
\end{align}
With the \( boost \) \( parameter \) \( \xi \) it said that
\begin{align}
       \beta &= \tanh \xi \\
      \gamma &= \cosh \xi \notag \\
\gamma \beta &= \sinh \xi \notag
\end{align}
applies to
\begin{align} \label{E:six}
x_0^{'} &= x_0 \cosh \xi - x_1 \sinh \xi \\
x_1^{'} &= -x_0 \sinh \xi + x_1 \cosh \xi. \notag
\end{align}

It is Einstein's thought that \( c = 1 \) with the velocity \( v \) 
in \( \tanh \xi = \frac{v}{c}. \) 
However, the wavespeed \( \lambda = \frac{f - 1}{f + 1} = \frac{1}{t} \)
and frequency \( f = \frac{t + 1}{t - 1} \)
relate differing wavespeeds with a differing number of seconds \( t. \)
Rather than using the boost parameter \( \xi \) in \eqref{E:six}, the hyperbolic coordinates 
are equated with the circular coordinates 
by \( \lambda = \tan \frac{\psi}{2} = \tanh \frac{a}{2}. \)  
The Lorentz time \( t \) equalling the distance over \( c \) of 
the \( x_{0}^{'} \) observer becomes
\begin{align}
x_{0}^{'} &= x_{0} \cosh \frac{a}{2} - x_{1} \sinh \frac{a}{2} = (\cosh \frac{a}{2})(x_{0} - x_{1} \tanh \frac{a}{2}) = (1 - \lambda^2)^{-1/2}(x_{0} - x_{1} \lambda) \\
x_{1}^{'} &= -x_{0} \sinh \frac{a}{2} + x_{1} \cosh \frac{a}{2} = (\cosh \frac{a}{2})(x_{1} - x_{0} \tanh \frac{a}{2}) = (1 - \lambda^2)^{-1/2}(x_{1} - x_{0}\lambda). \notag
\end{align}
The moving particle has advanced a distance \( v dt = \lambda^2 dt = d x_{0}^{'}. \)
Time is measured with twice the distance \( L \) of the hypotenuse,
vs. twice the height \( D \) of the side of the triangle.
The \( proper \) \( time \) observer sees \( x_{0} = 2D/\lambda. \)
The \( x_{0}^{'} \) observer, however, sees \( x_{0}^{'} = 2L/{\lambda}, \)
where \( L = ((\lambda^{2}x_{0}^{'}/2)^2 + D^2)^{1/2} \) and \( D = \lambda x_{0}/2, \)
by which 
\begin{equation}
x_{0}^{'} = (1 - \lambda^2)^{-1/2} x_{0}  
\end{equation}
is derivable when \( x_{1} = 0 \) occurs simultaneously.  
Moving clocks run slow.
Events will be separated by the time interval
\begin{equation}
x_{0}^{'} = (1 - \lambda^2)^{-1/2} \lambda x_{1} 
\end{equation}
since \( x_{0} = 0, \) although the events are simultaneous in time.

The element of arc length \( ds \) of the particle's path has \( ds^2 = c^2dt^2 - |dx|^2 \)
when \( dx_1 = r \sin \psi ds, \) \( dx_2 = \cos \psi ds, \) \( dx_3 = \sin \psi ds \)
are ``increments'' of \( x_1, x_2, x_3 \) having the angle \( \psi. \)
The direction of the path curve's polar-equation-tangent determined by the
angle \( \phi \) which this tangent makes with the radius \( r \) or by the 
angle \( \psi = \theta + \phi \) which it makes with the x-axis thereby
constructs \( dr = \cos \phi ds, \) \( rd\varphi = \sin \phi ds, \) and \( r \sin \varphi d\theta \)
\cite[pp. 120--121]{hC61}.
A motionless particle has \( ds^2 = dt^2. \) 
The new distance \( s \) is called \( proper \) \( time \) \( \tau, \)
and the \( Lorentz \) \( metric \) is
\( d\tau^{2} = c^{2}dt^{2}  - r^{2}\sin^{2}\varphi d\theta^{2} - dr^{2} - r^{2}d\varphi^{2}. \)
 
The square of the corresponding infinitesimal invariant interval \( ds \) is
\begin{align}
ds^2 &= c^{2}dt^2 - |dx|^2 \\ 
     &= c^{2}dt^{2}(1 - \beta^2) \notag \\ 
     &= \lambda^{2}dt^{2}(1 - \lambda^2) \notag
\end{align}
where \( \beta = \frac{v}{c} = \tanh a \) 
or where \( c \) is replaced by \( \lambda = \tanh \frac{a}{2} \)
and  \( v \) by \( \lambda^2 \).
In the coordinate system \( K' \) where the system is instantaneously at rest,
the space-time increments are \( dt' = d\tau, \) \( dx' = 0. \) 
Thus the invariant interval is \( ds = c d\tau\) or \(ds = \lambda d\tau \).
The increment of time \( d\tau \) in the instantaneous rest frame of the system
is an invariant quantity that takes the form
\begin{align}
d\tau &= dt(1 - \beta^2(t))^{1/2}  = \frac{dt}{\gamma(t)} \\
d\tau &= dt(1 - \lambda^2(t))^{1/2} \notag  
\end{align}
where \( \gamma = \cosh a = (1 - \beta^2)^{-1/2} \) 
and \( \cosh \frac{a}{2} = (1 - \lambda^2)^{-1/2}. \)
That is the time as seen in the rest frame of the system \cite[pp. 524--528]{jJ99}.

\section{FROM THE METRIC TO THE ANGLE OF PARALLELISM}
Further definition of the metric toward the \( physical \) \( angle \)
\( of \) \( parallelism, \) \( 2 \tan^{-1} e^{-a/s}, \) uses the arc length 
\begin{equation}
s = \int ds = \int_{x1}^{x2} (1 + (\frac{dy}{dx})^2)^{1/2} dx = \int_{t1}^{t2} ((\frac{dx}{dt})^2 + (\frac{dy}{dt})^2)^{1/2} dt. 
\end{equation}
Using the Gudermann 
\begin{align}
\psi &= \sin^{-1} \tanh a = \cos^{-1} \operatorname{sech} a = \tan^{-1} \sinh a \\
     &= \csc^{-1} \coth a = \sec^{-1} \cosh a = \cot^{-1} \operatorname{csch} a \notag
\end{align}
to solve the arc length of a unit circle from \( (0, 1) \) to \( (\cos \psi, \sin \psi), \)
that is, \( (\operatorname{sech} a, \tanh a), \) becomes
\begin{align}
s &= \int_{0}^{\operatorname{sech} a} (1 + (\frac{dy}{dx})^2)^{1/2} dx = \int_{0}^{a} ((\frac{dx}{dt})^2 + (\frac{dy}{dt})^2)^{1/2} dt \\
  &= \int_{0}^{a} ((-\operatorname{sech} t \tanh t)^2 + (\operatorname{sech}^2 t)^2)^{1/2} dt = \int_{0}^{a} \frac{2}{e^t + e^{-t}} dt \notag \\
  &= \int_{1}^{e^a} \frac{2}{u + \frac{1}{u}} \frac{du}{u} = 2 \tan^{-1} e^a - \frac{\pi}{2} = \psi. \notag
\end{align}
Solving the \( s \) arclength of a horizontal unit hyperbola \( x^2 - y^2 = 1 \)
from \( (1, 0) \) to \( (\cosh a, \sinh a) \) becomes
\begin{equation}
s = \int_{1}^{\cosh a} (1 + (\frac{dy}{dx})^2)^{1/2} dx = \int_{0}^{a} ((\frac{dx}{dt})^2 + (\frac{dy}{dt})^2)^{1/2} dt = \int_{0}^{a} (\sinh^2 t + \cosh^2 t)^{1/2} dt, 
\end{equation}
also known as the hyperbolic \( radius \) \( of \) \( curvature. \)

Lobacevskii's simpler angle of parallelism \cite[pp. 11--45]{nL91} \cite[pp. 376--377]{hC61} is
\begin{equation}
2 \tan^{-1} e^{-a} = 2 \tan^{-1} \frac{1}{f} = \frac{\pi}{2} - \psi = \frac{\pi}{2} - 2 \tan^{-1} \lambda = \theta,
\end{equation}
where \( 0 < \theta < \frac{\pi}{2}, \)
\begin{align}
\theta &= \sin^{-1} \operatorname{sech} a = \cos^{-1} \tanh a = \tan^{-1} \operatorname{csch} a \\
       &= \csc^{-1} \cosh a = \sec^{-1} \coth a = \cot^{-1} \sinh a \notag
\end{align}
is of a vertical unit hyperbola \( y^2 - x^2 = 1. \)
As \( \lambda \) approaches \( 1 \), \( 2 \tan^{-1} e^{-a} \) approaches \( 0 \).
With the hyperbolic radius of curvature \( s,\) however,
\( 2 \tan^{-1} e^{-a/s} \) approaches \( \frac{\pi}{2} \) radians, 
implying parallelism \cite[pp. 414, 434]{gM72} \cite[p. 217]{rE99} \cite[p. 315]{hC61}.
The \( metric \) is further defined to involve the physical angle of parallelism, 
the main conclusion of hyperbolic geometry \cite[pp. 216--218]{rE99} \cite[p. 77]{cM70} \cite[p. 210]{rM81}.
In the case of infinite seconds, the physical angle of parallelism 
is \( 2 \tan^{-1} e^{-1} = 0.705026844... \) radians, in agreement 
with the physical evidence cited by Martin \cite[p. 300]{gM72} \cite[p. 220]{rE99}.
The base \( a = \ln f \) of such astronomical asymptotic triangles is an 
important new kind of curve which is orthogonal to all parallels. 

\section{AN IMPROVED ELECTROMAGNETIC SPECTRUM}
The electromagnetic spectrum is illustrated with the Table, to be read with
all eight columns viewed on both facing pages.
The eight related categories, along with the electromagnetic nomenclature, are formulated

\begin{align} \label{E:four}
\mathrm{wavespeedmeters} &= \mathrm{wavelengthmeters \frac{meters}{cycle} \times frequency \frac{cycles}{second} = 299792458 \frac{meters}{second}} \notag \\                  
\mathrm{wavespeedradians} &= \lambda = \mathrm{wavelengthr \frac{radians}{cycle} \times frequency \frac{cycles}{second} = \frac{f - 1}{f + 1} = \frac{radians}{second}} \notag \\
                       &= \mathrm{wavelengthradians \frac{radians}{cycle} \times frequency \frac{1}{299792458} \frac{cycles}{meter} = \frac{radians}{meter}} \notag \\                                       
\mathrm{frequency} \quad f &= \mathrm{\frac{cycles}{second} = \frac{1}{299792458} \frac{cycles}{meter} = \frac{seconds + 1}{seconds - 1}} \notag \\                      
\mathrm{wavelengthradians} &= \mathrm{wavespeedradians \frac{radians}{second} \times \frac{1}{frequency} \frac{seconds}{cycle} = \frac{radians}{cycle}} \notag \\
                       &= \mathrm{wavespeedradians \frac{radians}{meter} \times \frac{1}{frequency} 299792458 \frac{meters}{cycle} = \frac{radians}{cycle}} \notag \\
\mathrm{wavelengthmeters} &= \mathrm{299792458 \frac{meters}{second} \times \frac{1}{frequency} \frac{seconds}{cycle} = \frac{meters}{cycle}} \notag \\
\mathrm{absolute acceleration} &= (1/2) \sec^{2} (\psi/2) \notag \\
\mathrm{relative acceleration} &= (1/2) \operatorname{sech}^{2} (a/2) = 2f/(f+1)^{2} \notag \\
\mathrm{instantaneousvelocityv} &= \lambda^2 \mathrm{\frac{radians}{second}} \notag \\
      \Pi(\frac{a}{s}) &= 2 \tan^{-1} e^{-a/s}, \quad a = \ln f, \quad s = \int_{0}^{a} ((\sinh t)^2 + (\cosh t)^2)^{1/2} dt. \notag  
\end{align}

A second with meters is singular, while seconds with radians increase from
one to infinity.
You see an infinite number of cycles apply to 1 second per radian,
a smaller 3 cycles apply to 2 seconds, 2 cycles apply to 3 seconds, and 1 cycle 
applies to an infinite number of seconds.
Thus an astronomer might observe a star to have a wavelength of 1.67 meters
per cycle, with a frequency of \( e^{19} \) cycles per second.
Reading the equivalent radian values on the same row of the Table,   
he could also know that there are \( 1.0 + (1.12 \times 10^{-8}) \) seconds per radian, 
a wavelength of \( 5.60 \times 10^{-9} \) radians per cycle, 
a wavespeed of \( \lambda = 1.0 - (1.12 \times 10^{-8}) \) radians per second,
absolute acceleration of \( 0.999999989, \)
relative acceleration of \( 0.000000011, \) 
and instantaneous velocity of \( 1.0 - (2.24 \times 10^{-8}) \) radians per second,
and an angle of parallelism 
of \( \Pi(\frac{a}{s}) = \frac{\pi}{2} - (1.50 \times 10^{-7}) \) radians per second \cite{rE89}.   

The additional radian information constructs a theory of time and geometry
to the universe.
All of the values are fixed throughout the Table. 
The Equivalence Principle is refined with values of the variable \( \lambda \) 
based upon the speed of light, \( c \) \cite[pp. 22--25]{wR69}.
In this manner, the universe may be measured mathematically \cite{sH88}.

\section{CONCLUSION}
George Martin (1932-- ) asks an interesting question \cite[pp. 300--302]{gM72},
\begin{quotation}
``A \( meter \) was originally intended to be one ten-millionth of the distance
from the earth's equator to a pole measured along a meridian. . . .
[It] turns out to be mathematically convenient to pick a [distance] scale
such that \( \Pi(1) \) is \( 2\arctan e^{-1}. \)
In applying this to the physical world, there is little difficulty in determining A, B, C
such that [angle] ABC has measure quite close to \( 2 \arctan e^{-1}. \)
However, how \( long \) is a segment of \( length \) 1?
That is, how many \( meters \) \( long \) is it?
Since a meter has nothing to do with the axioms of our geometry,
the question is a valid one.
Although it may be really neat to have a geometry that provides for a standard angle
determining a standard length,
all physical experiments indicate that \( \Pi(x) \) could noticeably differ from \( \frac{\pi}{2} \)
only for very large astromical distances \( x \).
So a physical segment of length 1 would be very, very \( long \) indeed.''
\end{quotation}
The answer is 299792458 meters long. Fewer meters apply to larger angles of parallelism.
Martin continues,
\begin{quotation}
``The largest physical triangles that can be accurately measured are astronomical.
Let \( E \) stand for the Earth, \( S \) for the Sun, and \( V \) for the 
brilliant blue star Vega. \( \measuredangle SEV \) can be measured from the Earth when
\( \measuredangle ESV \) is right. Using this measurement and the fact that [the
defect of the triangle] \( SEV \) is less than \( \frac{\pi}{2} \) -
\( \measuredangle SEV, \) one obtains [the defect of the triangle] \( SEV < 0.0000004. \) ''
\end{quotation}   
In our Table, that agrees closely to a Short-wave radio (blue) 
angle of parallelism.
By comparison, one \( parsec \) is defined as the distance of the Earth
to the Sun (1 astronomical unit) that subtends an angle of 1 second of arc,
equivalent to 206265 astronomical units.
The comparable \( ES \) base of the asymptotic triangle 
is about \( a = \ln e^{19} = 19 \) radial units.
With the angle of parallelism and electromagnetic spectrum,
meters and seconds can hereby be exchanged with radians and seconds. 
We can geometrically measure the universe. 

\newpage
\begin{center}
\begin{tabular}{||c|c|c|c|c|}
\hline
seconds/radian                              & frequency                               & wavelength r                & wavelength m               & absolute acceleration \\
\( 1 / \lambda \)                           & \( cycles/second \)                     & \( radians/cycle \)         & \( meters/cycle \)         & \( d\lambda/d\psi \) \\ \hline
\( 1.0 \)                                   & \( \infty \)                            & \( 0 \)                     & \( 0 \)                    & \( 1.0 \) \\
\( 1.0 + (1.92 \times 10^{-23}) \)          & \( e^{53} = 1.04 \times 10^{23} \)      & \( 9.60 \times 10^{-24} \)  & \( 2.87 \times 10^{-15} \) & \( 1.0 - (1.92 \times 10^{-23}) \) \\  
\( 1.0 + (1.41 \times 10^{-22}) \)          & \( e^{51} = 1.40 \times 10^{22} \)      & \( 7.09 \times 10^{-23} \)  & \( 2.12 \times 10^{-14} \) & \( 1.0 - (1.41 \times 10^{-22}) \) \\ 
\( 1.0 + (1.04 \times 10^{-21}) \)          & \( e^{49} = 1.90 \times 10^{21} \)      & \( 5.24 \times 10^{-22} \)  & \( 1.57 \times 10^{-13} \) & \( 1.0 - (1.04 \times 10^{-21}) \) \\
\( 1.0 + (7.74 \times 10^{-21}) \)          & \( e^{47} = 2.58 \times 10^{20} \)      & \( 3.87 \times 10^{-21} \)  & \( 1.16 \times 10^{-12} \) & \( 1.0 - (7.74 \times 10^{-21}) \) \\                       
\( 1.0 + (5.72 \times 10^{-20}) \)          & \( e^{45} = 3.49 \times 10^{19} \)      & \( 2.86 \times 10^{-20} \)  & \( 5.74 \times 10^{-12} \) & \( 1.0 - (5.72 \times 10^{-20}) \) \\  
\( 1.0 + (4.23 \times 10^{-19}) \)          & \( e^{43} = 4.72 \times 10^{18} \)      & \( 2.11 \times 10^{-19} \)  & \( 6.34 \times 10^{-11} \) & \( 1.0 - (4.23 \times 10^{-19}) \) \\
\( 1.0 + (3.12 \times 10^{-18}) \)          & \( e^{41} = 6.39 \times 10^{17} \)      & \( 1.56 \times 10^{-18} \)  & \( 4.68 \times 10^{-10} \) & \( 1.0 - (3.12 \times 10^{-18}) \) \\  
\( 1.0 + (2.30 \times 10^{-17}) \)          & \( e^{39} = 8.65 \times 10^{16} \)      & \( 2.30 \times 10^{-17} \)  & \( 3.46 \times 10^{-9} \)  & \( 1.0 - (2.30 \times 10^{-17}) \) \\
\( 1.0 + (1.70 \times 10^{-16}) \)          & \( e^{37} = 1.17 \times 10^{16} \)      & \( 8.53 \times 10^{-17} \)  & \( 1.71 \times 10^{-8} \)  & \( 1.0 - (1.70 \times 10^{-16}) \) \\
\( 1.0 + (1.26 \times 10^{-15}) \)          & \( e^{35} = 1.58 \times 10^{15} \)      & \( 6.30 \times 10^{-16} \)  & \( 1.89 \times 10^{-7} \)  & \( 1.0 - (1.26 \times 10^{-15}) \) \\
\( 1.0 + (9.31 \times 10^{-15}) \)          & \( e^{33} = 2.14 \times 10^{14} \)      & \( 4.65 \times 10^{-15} \)  & \( 1.39 \times 10^{-6} \)  & \( 1.0 - (9.31 \times 10^{-15}) \) \\
\( 1.0 + (6.88 \times 10^{-14}) \)          & \( e^{31} = 2.90 \times 10^{13} \)      & \( 3.44 \times 10^{-14} \)  & \( 0.0000103 \)            & \( 1.0 - (6.88 \times 10^{-14}) \) \\
\( 1.0 + (5.08 \times 10^{-13}) \)          & \( e^{29} = 3.93 \times 10^{12} \)      & \( 2.54 \times 10^{-13} \)  & \( 0.0000762 \)            & \( 1.0 - (5.08 \times 10^{-13}) \) \\
\( 1.0 + (3.75 \times 10^{-12}) \)          & \( e^{27} = 5.32 \times 10^{11} \)      & \( 1.87 \times 10^{-12} \)  & \( 0.0005634 \)            & \( 1.0 - (3.75 \times 10^{-12}) \) \\                          
\( 1.0 + (2.77 \times 10^{-11}) \)          & \( e^{25} = 7.20 \times 10^{10} \)      & \( 1.38 \times 10^{-11} \)  & \( 0.0041635 \)            & \( 1.0 - (2.77 \times 10^{-11}) \) \\
\( 1.0 + (2.05 \times 10^{-10}) \)          & \( e^{23} = 9.74 \times 10^{9} \)       & \( 1.02 \times 10^{-10} \)  & \( 0.0307643 \)            & \( 1.0 - (2.05 \times 10^{-10}) \) \\
\( 1.0 + (1.51 \times 10^{-9}) \)           & \( e^{21} = 1.31 \times 10^{9} \)       & \( 7.58 \times 10^{-10} \)  & \( 0.2273194 \)            & \( 0.999999998 \) \\                        
\( 1.0 + (6.67 \times 10^{-9}) \)           & \( 299792458 \)                         & \( 3.33 \times 10^{-9} \)   & \( 1.0 \)                  & \( 0.999999993 \) \\
\( 1.0 + (1.12 \times 10^{-8}) \)           & \( e^{19} = 1.78 \times 10^{8} \)       & \( 5.60 \times 10^{-9} \)   & \( 1.6796761 \)            & \( 0.999999989 \) \\
\( 1.0 + (8.27 \times 10^{-8}) \)           & \( e^{17} = 2.41 \times 10^{7} \)       & \( 4.13 \times 10^{-8} \)   & \( 12.411221 \)            & \( 0.999999917 \) \\
\( 1.0 + (6.11 \times 10^{-7}) \)           & \( e^{15} = 3.26 \times 10^{6} \)       & \( 3.05 \times 10^{-7} \)   & \( 91.707208 \)            & \( 0.999999388 \) \\
\( 1.0 + (4.52 \times 10^{-6}) \)           & \( e^{13} = 442413.39 \)                & \( 2.26 \times 10^{-6} \)   & \( 677.62970 \)            & \( 0.999995479 \) \\
\( 1.000033403 \)                           & \( e^{11} = 59874.141 \)                & \( 0.00001670 \)            & \( 5007.0439 \)            & \( 0.999966598 \) \\
\( 1.000256850 \)                           & \( e^{9} = 8103.0839 \)                 & \( 0.00012337 \)            & \( 36997.328 \)            & \( 0.999753241 \) \\
\( 1.001825428 \)                           & \( e^{7} = 1096.6331 \)                 & \( 0.00091022 \)            & \( 273375.33 \)            & \( 0.998179558 \) \\
\( 1.013567309 \)                           & \( e^{5} = 148.41315 \)                 & \( 0.00664775 \)            & \( 2.01 \times 10^{6} \)   & \( 0.986703887 \) \\
\( 1.104791392 \)                           & \( e^{3} = 20.085536 \)                 & \( 0.04506467 \)            & \( 1.49 \times 10^{7} \)   & \( 0.909646681 \) \\
\( 1.313035285 \)                           & \( e^{2} = 7.3890560 \)                 & \( 0.10307056 \)            & \( 4.05 \times 10^{7} \)   & \( 0.790012829 \) \\
\( 2 \)                                     & \( 3/1 \)                               & \( 1/(2 \times 3) \)        & \( 9.99 \times 10^{7} \)   & \( 5/8 \) \\
\( 2.163953413 \)                           & \( e^{1} = 2.7182818 \)                 & \( 0.17000340 \)            & \( 1.10 \times 10^{8} \)   & \( 0.606776134 \) \\
\( 3 \)                                     & \( 4/2 \)                               & \( 2/(3 \times 4) \)        & \( 1.49 \times 10^{8} \)   & \( 5/9 \) \\
\( 4 \)                                     & \( 5/3 \)                               & \( 3/(4 \times 5) \)        & \( 1.79 \times 10^{8} \)   & \( 17/32 \) \\
\( 5 \)                                     & \( 6/4 \)                               & \( 4/(5 \times 6) \)        & \( 1.99 \times 10^{8} \)   & \( 13/25 \) \\
\( \infty \)                                & \( 1.0 \)                               & \( 1/\infty \)              & \( 299792458 \)            & \( 1/2 \) \\ \hline
\end{tabular}
\end{center}

\newpage

\begin{center}
\begin{tabular}{|c|c|c|l||}
\hline
relative accel                                   & transport acceleration                    & angle of parallelism                          & \\
\( d\lambda/da \)                                & instant velocity \( \lambda^2 \)           & \( \Pi(\frac{a}{s}) = 2 \tan^{-1} e^{-a/s} \) & Nomenclature \\ \hline 
\( 0 \)                                          & \( 1.0 \)                                 & \( \pi/2 \)                                   & \\
\( 1.92 \times 10^{-23} \)                       & \( 1.0 - (3.84 \times 10^{-23}) \)        & \( \pi/2 - (7.19 \times 10^{-22}) \)          & Cosmic photons \\
\( 1.41 \times 10^{-22} \)                       & \( 1.0 - (2.83 \times 10^{-22}) \)        & \( \pi/2 - (5.11 \times 10^{-21}) \)          & \(\gamma\) - rays \\
\( 1.04 \times 10^{-21} \)                       & \( 1.0 - (2.09 \times 10^{-21}) \)        & \( \pi/2 - (3.63 \times 10^{-20}) \)          & X-rays\\
\( 7.74 \times 10^{-21} \)                       & \( 1.0 - (1.54 \times 10^{-20}) \)        & \( \pi/2 - (2.57 \times 10^{-19}) \)          & \\
\( 5.72 \times 10^{-20} \)                       & \( 1.0 - (1.14 \times 10^{-19}) \)        & \( \pi/2 - (1.82 \times 10^{-18}) \)          & \\
\( 4.23 \times 10^{-19} \)                       & \( 1.0 - (8.46 \times 10^{-19}) \)        & \( \pi/2 - (1.28 \times 10^{-17}) \)          & Soft x-rays \\
\( 3.12 \times 10^{-18} \)                       & \( 1.0 - (6.25 \times 10^{-18}) \)        & \( \pi/2 - (9.06 \times 10^{-17}) \)          & Ultraviolet\\
\( 2.30 \times 10^{-17} \)                       & \( 1.0 - (4.61 \times 10^{-17}) \)        & \( \pi/2 - (6.36 \times 10^{-16}) \)          & \\
\( 1.70 \times 10^{-16} \)                       & \( 1.0 - (3.41 \times 10^{-16}) \)        & \( \pi/2 - (4.46 \times 10^{-15}) \)          & Visible spectrum\\
\( 1.26 \times 10^{-15} \)                       & \( 1.0 - (2.52 \times 10^{-15}) \)        & \( \pi/2 - (3.12 \times 10^{-14}) \)          & Infrared\\
\( 9.31 \times 10^{-15} \)                       & \( 1.0 - (1.86 \times 10^{-14}) \)        & \( \pi/2 - (2.17 \times 10^{-13}) \)          & \\
\( 6.88 \times 10^{-14} \)                       & \( 1.0 - (1.37 \times 10^{-13}) \)        & \( \pi/2 - (1.50 \times 10^{-12}) \)          & \\
\( 5.08 \times 10^{-13} \)                       & \( 1.0 - (1.01 \times 10^{-12}) \)        & \( \pi/2 - (1.04 \times 10^{-11}) \)          & Far-infrared \\
\( 3.75 \times 10^{-12} \)                       & \( 1.0 - (7.51 \times 10^{-12}) \)        & \( \pi/2 - (7.17 \times 10^{-11}) \)          & Microwaves \\
\( 2.77 \times 10^{-11} \)                       & \( 1.0 - (5.55 \times 10^{-11}) \)        & \( \pi/2 - (4.19 \times 10^{-10}) \)          & \\
\( 2.05 \times 10^{-10} \)                       & \( 1.0 - (4.10 \times 10^{-10}) \)        & \( \pi/2 - (3.33 \times 10^{-9}) \)           & \\
\( 0.000000002 \)                                & \( 1.0 - (3.03 \times 10^{-9}) \)         & \( \pi/2 - (2.25 \times 10^{-8}) \)           & Television \\
\( 0.000000007 \)                                & \( 1.0 - (1.33 \times 10^{-8}) \)         & \( \pi/2 - (9.20 \times 10^{-8}) \)           & FM radio \\
\( 0.000000011 \)                                & \( 1.0 - (2.24 \times 10^{-8}) \)         & \( \pi/2 - (1.50 \times 10^{-7}) \)           & Short-wave radio\\
\( 0.000000083 \)                                & \( 1.0 - (1.65 \times 10^{-7}) \)         & \( \pi/2 - (9.95 \times 10^{-7}) \)           & AM radio \\
\( 0.000000612 \)                                & \( 1.0 - (1.22 \times 10^{-6}) \)         & \( \pi/2 - (6.48 \times 10^{-6}) \)           & Long-wave radio\\
\( 0.000004521 \)                                & \( 1.0 - (9.04 \times 10^{-6}) \)         & \( 1.57075477 \)                              & Induction heating\\
\( 0.000033402 \)                                & \( 0.99993319 \)                          & \( 1.57053650 \)                              & \\
\( 0.000246759 \)                                & \( 0.99950648 \)                          & \( 1.56922541 \)                              & \\                               
\( 0.001820442 \)                                & \( 0.99635911 \)                          & \( 1.56176230 \)                              & Power \\
\( 0.013296113 \)                                & \( 0.97340777 \)                          & \( 1.52289663 \)                              & \\         
\( 0.090353319 \)                                & \( 0.81929336 \)                          & \( 1.35203152 \)                              & \\
\( 0.209987171 \)                                & \( 0.58002565 \)                          & \( 1.15128526 \)                              & \\
\( 3/8 \)                                        & \( 1/4 \)                                 & \( 0.90301878 \)                              & \\
\( 0.393223866 \)                                & \( 0.21355226 \)                          & \( 0.87551570 \)                              & \\
\( 4/9 \)                                        & \( 1/9 \)                                 & \( 0.79641144 \)                              & \\
\( 15/32 \)                                      & \( 1/16 \)                                & \( 0.75740822 \)                              & \\
\( 12/25 \)                                      & \( 1/25 \)                                & \( 0.73888970 \)                              & \\
\( 1/2 \)                                        & \( 1/\infty \)                            & \( 0.70502684 \)                              & \\ \hline                                                               
\end{tabular}
\end{center}
\footnote{In the January 2001 Notices of the AMS, Irving Ezra Segal believed that
a static universe recessional velocity of \( \tan^2 \frac{\rho}{2}, \) \( \rho \)
radians, fit the evidence better than Hubble's linear proportionality.}

\newpage

\bibliographystyle{amsplain}
\bibliography{standardizing}

\end{document}

%% file: timemacros.tex
\renewcommand{\rm}{\normalshape}%

\newif\ifShowLabels
\ShowLabelstrue
\newdimen\theight
\def\TeXref#1{%
     \leavevmode\vadjust{\setbox0=\hbox{{\tt
          \quad\quad  {\small \rm #1}}}%
     \theight=\ht0
     \advance\theight by \dp0
     \advance\theight by \lineskip
     \kern -\theight \vbox to
     \theight{\rightline{\rlap{\box0}}%
      \vss}%
      }}%

     { \begin{Thm} \label{T:#1} \ifShowLabels \TeXref{T:#1} \fi }%
     { \end{Thm} }

     { \begin{Def} \label{D:#1} \ifShowLabels \TeXref{D:#1} \fi }%
     { \end{Def} }

     { \begin{Exe} \label{X:#1} \ifShowLabels \TeXref{X:#1} \fi }%
     { \end{Exe} }

\newcommand{\EqRef}[1]%
     { \ifShowLabels \TeXref{E:#1} \fi
       \begin{equation} \label{E:#1} }

    {\par \samepage}%
    {\par}
 

%% file: standardizing.bbl
\ifx\undefined\bysame
\newcommand{\bysame}{\leavevmode\hbox to3em{\hrulefill}\,}
\fi
\begin{thebibliography}{10}

\bibitem{gC31}
G.~Chrystal, {\em Algebra}, vol.~II, Black, London, 1931.

\bibitem{hC61}
H.~S.~M. Coxeter, {\em Introduction to geometry}, 4th ed., Toronto Univ. Press,
  Toronto, 1961.

\bibitem{aE57}
A.~Einstein, {\em Relativity, the special and the general theory}, 15th ed.,
  Methuen, London, 1957.

\bibitem{rE89}
R.~C. Eskew, {\em Relating a circle and a hyperbola}, J. Undergrad. Math. {\bf
  21(2)} (1989), 49--54.

\bibitem{rE99}
\bysame, {\em Time, gravity and the exterior angle of parallelism},
  Speculations in Science and Technology {\bf 21} (1999), 213--225.

\bibitem{sH88}
S.~W. Hawking, {\em A brief history of time}, Bantam, Toronto, 1988.

\bibitem{iI69}
I.~M. Iaglom, {\em A simple non-euclidean geometry and its physical basis},
  Nauka, Moscow, 1969 (Russian), English translation availble.

\bibitem{jJ99}
J.~D. Jackson, {\em Classical electrodynamics}, 3rd ed., Wiley, New York, 1999.

\bibitem{nL91}
N.~I. Lobachevskii, {\em Geometrical researches on the theory of parallels},
  Karzan, Berlin, 1840 (Russian), English translation available.

\bibitem{hL23}
H.~A. Lorentz, {\em The principle of relativity}, Dover, New York, 1923.

\bibitem{gM72}
G.~E. Martin, {\em The foundations of geometry and the non-euclidean plane},
  Intext Educational, New York, 1972.

\bibitem{rM81}
R.~S. Millman, {\em Geometry, a metric approach}, Springer-Verlag, New York,
  1981.

\bibitem{cM70}
C.~W. Misner, {\em Gravitation}, Freeman, San Francisco, 1970.

\bibitem{wR69}
W.~Rindler, {\em Essential relativity: special, general, and cosmological}, 2nd
  ed., Van Nostrand Reinhold, New York, 1969.

\end{thebibliography}
